\documentstyle[12pt]{article}
\begin{document}

\newcommand{\be}{\begin{equation}}
\newcommand{\ee}{\end{equation}}
\newcommand{\ba}{\begin{eqnarray}}
\newcommand{\ea}{\end{eqnarray}}

\begin{center}
{\bf N.E.Tyurin}\\

\vspace*{0.2cm}
{\bf SOME ASPECTS OF SPIN PHYSICS AT RHIC}\\

\vspace*{0.2cm}
(\bf Talk presented at PHENIX project meeting. \\
Vladimir, Russia, November 9-11, 1995)\\

\end{center}

\vspace*{0.3cm}

\section*{Introduction}
Spin physics is becoming a very popular topic during last few
years. Now it is generally accepted that spin plays an important role
in particle and high energy physics~[1-3].
It provides crucial  information on such
fundamental issues as
\begin{itemize}
\item hadron wave functions
\item interplay of large and short distance dynamics
\item chiral symmetry breaking and confinement.
\end{itemize}

To proceed in our
understanding of the field we should answer the questions:
\begin{itemize}
\item How the constituents' interactions depend on spin?
\item What is a picture of the proton structure, in particular,
its spin structure?
\end{itemize}

Spin experiments at RHIC are planned to provide systematic
studies of the above issues. Spin program has been approved for
both machine  and detectors and RHIC operation before year~2000
is a realistic goal. All that cause great interest and
enthusiasm of the physics community. The subject was extensively
discussed in June, 1995 and could be found in distributed set of
transparencies entitled ``Review of the RHIC Spin Physics Program''.

In this talk we consider some aspects~[1,4] which have not
received much attention in the above mentioned review. In
particular, we discuss spin properties of QCD, basic
experimental results which continue to be an engine of this
topic, physics to be done at RHIC with particular emphasis on
asymmetries in inclusive hadron production.

\section*{Spin Properties of QCD}
The QCD Lagrangian has the form
\ba
L_{QCD}&=& \bar\psi (x)(i\gamma^\mu D_\mu - \hat m)
\psi (x)-\frac{1}{4}\mbox{tr} (G_{\mu\nu}G^{\mu\nu}),\\
D_\mu&=&\partial_\mu -ig\frac{\lambda^a}{2} G^a_\mu,\,\,
\hat m=\mbox{diag}\,(m_u, m_j, m_s).\nonumber
\ea
$G^a_{\mu\nu}$ is the gluon field strength tensor, $\lambda^a$
are the generators of the $SU(3)$ color group. It describes
interactions between quarks and gluons --- the hadron
constituents. Contrary to QED, this Lagrangian describes
self-interaction of the massless color gluons. The theory is
nonlinear over the gauge fields $G^a_\mu$, and the coupling
constant is large. Fortunately because of asymptotic freedom~[5]
the running coupling constant $\alpha_s(Q^2)$ becomes small at
the scale $Q^2>\Lambda^2_{QCD}$. The parameter
$(\Lambda_{QCD})^{-1}$ determines the scale of short distances.
Thus one introduces perturbative and non-perturbative ``phases''
of QCD.

The current quark masses in $L_{QCD}$ are small. In the limit
$m_q\to 0$ $L_{QCD}$ is invariant under the chiral group
$SU(3)_L\times SU(3)_R$. Chiral invariance and vector nature of
QCD impose important constraints on spin observables. In the
chiral limit $m_q\to 0$ the QCD interactions are the same for
the left and right quarks
$$
\bar\psi\hat D\psi=\bar\psi_L\hat D\psi_L
+\bar\psi_R\hat D\psi_R,
$$
$$
\psi_L=\frac{1}{2}(1-\gamma_5)\psi,\,\,
\psi_R=\frac{1}{2}(1+\gamma_5)\psi,
$$
$$
\gamma_5\psi_{L,R}=\mp\psi_{L,R}.
$$
For massless quarks chirality and helicity coincide
$$
\psi_{1/2}=\psi_R,\,\, \psi_{-1/2}=\psi_L.
$$
The quark helicity conservation is the most characteristic
feature of perturbative theory with vector coupling.

Because of a small mass in $L_{QCD}$
$$
\psi_{\pm 1/2}=\psi_{\stackrel{R}{L}}+0\left
(\frac{m}{\sqrt{\hat s}}\right ).
$$
In hard interactions polarization has to be vanishingly small
\be
P_q\propto \frac{\alpha_s m_q}{\sqrt{\hat s}},
\ee
where $\alpha_s$ and $m_q$ are small and $\sqrt{\hat s}\sim
p_\perp$ is large.

In fact the chiral group $SU(3)_L\times SU(3)_R$ is not realized
in the hadron spectrum and chirality is not a symmetry of QCD.
It is broken by the vacuum state $<\bar\psi\psi>\not= 0$. This
is a non-perturbative effect.

Perturbative QCD ($p$QCD) deals with perturbative vacuum
invariant under the chiral transformations. Nonperturbative
phase of QCD should provide the spontaneous breaking of chiral
symmetry. The relevant scale is characterized  by the parameter $\Lambda_\chi$.
The values of the parameters related to confinement and
spontaneous breaking of chiral symmetry are
$\Lambda_{QCD}=100\div 300$MeV and $\Lambda_\chi\simeq 1$GeV~[6].

At small distances $r<\Lambda_\chi^{-1}$ helicity is conserved
due to chiral invariance. At $r>\Lambda_{QCD}^{-1}$ there is no
need for helicity  conservation, but it appears
that asymmetries in this range are
close to zero.
In the range between the two scales chiral symmetry is
spontaneously broken and non-perturbative effects should be
taken into account although $\alpha_s$ is small at $r<\Lambda_{QCD}^{-1}$.

How to deal with non-perturbative phenomena? To use the ideas of
effective  Lagrangians' approach which was worked out in
details, for instance, Nambu-Jona-Lasinio model~[7]. The chiral
symmetry breaking results in particular in generation of quark
masses and in appearance of quark condensates:
$m_q\propto -\Lambda_\chi^{-2}<\bar\psi\psi>$. As a consequence
it assumes some constraints on hadron structure. We return to
this important issue in last part of this talk.

The basic tool for obtaining QCD predictions for hard processes
is the factorization which separates long distance dynamics of
the bound state system and short distance constituents' interaction.

The hadron wave function
$$
\psi^h_\lambda (x_i,\vec k_{\perp i}, \lambda_i),
$$
where $x_i$ is light-cone momentum fraction of the i-th
constituent (quark or gluon),
$\vec k_{\perp i}$ is its transverse momentum and $\lambda_i$
--- the helicity. It describes bound state of the constituents
and may be decomposed over Fock basis
$$
|p>=|qqq>\psi^p_{qqq}+|qqqg>\psi^p_{qqqg}+...
$$
For exclusive process $A+B\to C+D$ the hadron states with
constituents additional to valence quarks are suppressed. The
respective transition amplitude of a hadron~[8]
\be
\Phi^h_\lambda (x_i,\lambda_i;Q)\propto
\int^{k^2_{\perp i}<Q^2}
[d^2k_\perp]\psi^h_\lambda
(x_i,\vec k_{\perp i}, \lambda_i),
\ee
corresponds to representation of the hadron as a set of
collinear partons. Fig.1 shows the three valence quarks. As it
was said the other states are suppressed by the powers of $(Q^2)^{-1}$.

According the factorization theorem the hadron amplitude $T$ can
be presented in the form (Fig.2):
$$
T\propto \prod_{k=A,B,C,D} \Phi^k(x_i,\lambda_i, Q)\otimes
T_H(x_i, \lambda_i;Q)\;.
$$
This equation assumes integrations over momentum fractions for all
four hadrons as well as sum over $\lambda_i$. All the quarks are
in the $s$-state because of integration over $[d^2k_\perp]$ in
eq.(3) and consequently the hadron helicity
$$
\lambda_h=\sum\lambda_i.
$$
The hard amplitude $T_H$ conserves the quark helicity and we get
$$
\lambda_A+\lambda_B=\lambda_C+\lambda_D.
$$
This equality leads to very important experimental consequences
--- vanishing of the\\ one-spin asymmetries for hard exclusive processes:
$$
A_N=0.
$$
For inclusive process $A+B\to C+X$ the key instruments are
structure functions:
\ba
f_{a/h}(x,\lambda,Q)\propto
\sum_{n,\lambda_i}&\int& [dx][d^2k_\perp]\left |
\psi^h_n(x_i, \vec k_{\perp i}, \lambda_i)\right
|^2\cdot\nonumber \\
\cdot \sum_{b=a}&\delta& (x_b-x)\;.
\ea
Contrary to exclusive reaction all of the Fock states with
orbitrary numbers of quarks and gluons and values of orbital
angular momentum contribute
$$
\lambda_h\not=\sum_{i}\lambda_i.
$$
Therefore we should write
\be
\lambda_h=\lambda_q+\lambda_g+<L_z>_q+<L_z>_g.
\ee
This is the most general equation for a hadron helicity
distributed among its constituents.

If the structure functions eq.(4) are known the factorization
allows one to calculate the cross-sections and asymmetries. As
an illustration we give the two formulas (see fig.3) for the
asymmetries in inclusive $pp\to hX$ process:
\ba
A_N\sigma (pp\to hX)=
\sum_{a,b,c,d}&\int& dx_adx_b\frac{dz}{z}\Delta_\perp
f_{a/p}(x_a)\cdot f_{b/p}(x_b)\times \nonumber \\
&\times&
a_N\sigma (ab\to cd)\cdot D_{h/c}(z)\;,
\ea
\ba
A_{LL}\sigma (pp\to hX)=
\sum_{a,b,c,d}&\int& dx_a dx_b\frac{dz}{z}\Delta_L
f_{a/p}(x_a)\Delta_L f_{b/p}(x_b)\times \nonumber \\
&\times&
a_{LL}\sigma (ab\to cd)\cdot D_{h/c}(z)\;,
\ea
where notations were earlier introduced or are obvious. Each
factor $f_{a/h}$ is related to a single incoming hadron. It
makes useful the DIS data. Because of vanishing asymmetry in the
hard subprocess
$$
a\propto \frac{\alpha_s m_q}{\sqrt{s'}},\,\,\,\,
\sqrt{s'} \sim p_\perp,
$$
the one-spin transverse asymmetry $A_N$ at hadron level should
also vanish. This is true even if the above simple factorization
does not  work. The conclusion is that perturbative QCD predicts
$A_N=0$.

Apart of that factorization provides a powerfull and transparent
approach based on $L_{QCD}$ Eq.(1) to make calculations and
predictions of the asymmetries to be then experimentally measured.

What are the limitations of such calculations? The approach
assumes a parton picture of hadron consisting of free quarks and
gluons. In fact only interactions with $Q^2>\Lambda^2_\chi$ resolve
the partonic structure. The range $\Lambda^2_{QCD}<Q^2<\Lambda^2_\chi$
requires more care.

To rely on factorization predictions, for instance, under the
study of two-spin asymmetries $A_{LL}$ and $A_{NN}$, one should
ask the question: are we in the perturbative QCD sector? The
best test is to measure $A_N$ and to prove that $A_N=0$. It is a
great challenge to find a ``threshold'' $p_\perp$ value where
$p$QCD and factorization could be well justified. In the case
$A_N\not= 0$ one should estimate the potential sources of the
observed asymmetry. The present day experiment shows that
non-perturbative QCD effects are to be taken into account.

\section*{Basic Experimental Results}
1. It is well known that single-spin effects~[9]
in general do not
consent with $p$QCD. Fig.4 shows the asymmetry $A$ plotted
against $p^2_\perp$ for polarized $pp$ elastic scattering at 24
and 28~GeV/c. The asymmetry $A$ grows up with $p^2_\perp$
contrary to what one could expect from the perturbation theory.

The other important set of data [10] is related to $\Lambda$ ---
polarization in the inclusive production process $pp\to \Lambda
X$. In fig.5 the polarization of
$\Lambda$ hyperons produced by 400~GeV protons is plotted
against $p_\perp$. Fig.6 shows the data at 12~GeV and 2000~GeV
compared with the curve (dashed) which corresponds to 400~GeV
data. We conclude that polarization of $\Lambda$ does not depend
much on energy. It grows up and becomes constant at
$p_\perp \simeq 1$GeV/c. It is $x$-dependent function.

Analysis of the one-spin asymmetries data  leads to conclusion
on necessity of further experimental studies. If these effects
persist at higher energies it will strongly indicate on
non-perturbative origin of spin dynamics.

2. Measurements of polarized structure functions in DIS
processes~[11] revealed that quark contribution to the nucleon
spin is small. The other possibilities could be contribution  of
gluons and account for orbital angular momentum. Then the only
constraint is
\ba
\frac{1}{2}=\frac{1}{2}\Delta q&+&\Delta g+<L_z>_q+
<L_z>_g,\nonumber \\
\Delta q&=&\Delta u+\Delta d+\Delta s\;.
\ea
Knowing different terms in Eq.(8) allows one to approach the
picture of hadron structure and to clarify the role of
non-perturbative effects. Experiment shows that
\be
(\Delta u+\Delta d+\Delta s)_p\simeq 1/3,\,\,
\Delta s \simeq -0.1\;.
\ee
Therefore about $2/3$ of the proton spin is to be attributed to
orbital angular momentum and gluons. The role of $s$-quarks is
important. Probably one should say in general that role of
$\bar qq$-pairs in the proton structure is to be carefully traced.

\section*{The Structure Functions}
We introduce first distribution functions for quarks
$q_\lambda (x, Q^2)=f_{q/h}(x,\lambda;Q)$, antiquarks
$\bar q_\lambda (x,Q^2)$ and gluons $G_\lambda(x,Q^2)$. Then spin
dependent quark distributions are as follows:
\be
\Delta_L q(x, Q^2)=q_{\rightarrow} (x,Q^2)-
q_{\leftarrow} (x,Q^2)\;,
\ee
for longitudinally polarized proton and
\be
\Delta_\perp q(x,Q^2)=q_{\uparrow}(x,Q^2)-q_{\downarrow}(x,Q^2)\;,
\ee
for transversely polarized proton. For gluons
\be
\Delta G (x,Q^2)=G_{\rightarrow}(x, Q^2)-
G_{\leftarrow}(x, Q^2)\;.
\ee

Averaged over spin structure function
\be
f_1(x,Q^2)=\frac{1}{2}\sum_{q}
e^2_q [q(x,Q^2)+\bar q(x,Q^2)]\;.
\ee
Sometimes it is represented as
$$
f_1=f^q_1+f^{\bar q}_1.
$$
Spin structure function
\be
g_1(x,Q^2)=g_L(x,Q^2)=\frac{1}{2}\sum_{q}e^2_q
(\Delta_Lq+\Delta_L\bar q)
\ee
measures quark helicity distribution in a longitudinally
polarized nucleon. The function
\be
h_1(x,Q^2)=g_\perp (x,Q^2)=
\frac{1}{2}\sum_{q} e^2_q
\left (\frac{m_q}{xM}\right )\cdot
(\Delta_\perp q+\Delta_\perp \bar q)
\ee
measures the average transverse spin for quarks. In the simple
parton model with noninteracting partons, i.e. when hadron is
treated as a gas of free quarks, one should expect [12]
$$
\Delta_\perp q=\Delta_L q.
$$
The second spin dependent structure function is determined as
\be
g_2(x,Q^2)=h_1(x,Q^2)-g_1(x,Q^2)
\ee
and is related to effects of quark-gluon interactions or to the
higher twists.

The functions introduced satisfy to series of important
constraints and relations (sum rules)~[13]. They are often used
as a tool for QCD tests. Up to now only the longitudinal
spin-dependent structure functions of the nucleon were measured experimentally.

\section*{Spin Physics at RHIC}
The topic was extensively reviewed during dedicated meeting at
BNL~[14]. This paragraph contains only some of the relevant
directions (cf. [14,4,15]).

1. It is expected that \underline{enigma of the proton spin distribution}
between various constituents should be experimentally solved at
RHIC by measuring the separate contributions of gluons, valence
and sea quarks and angular momentum fraction.

We enlist below the reaction, fundamental subprocess (cf. fig3),
the parameter to be measured and the function which will be extracted:
\begin{center}
\begin{tabular}{lllll}
$\bullet\; \vec p+\vec p\to$ &
$\gamma+X\;\;\;$,&
$g +q\to \gamma +q$, & $A_{LL}\Rightarrow$ & $\Delta G$\\
 &$\gamma+\mbox{jet}$ & & & $\Delta G(x)$\\
 &$\mbox{jet}+X$ & & & \\[0.3cm]
$\bullet\; \vec p+ p\to$ &
$W^{+}+X\;\;\;$,&
$u +\bar d\to W^{+}$, & $A_L\;\;\Rightarrow $ & $\Delta \bar
d(x)/\bar d(x)$\\
 & & & & $\Delta u(x)/u(x)$\\[0.4cm]
$\bullet\; \vec p+ p\to$ &
$W^{-}+X\;\;\;$,&
$\bar u + d\to W^{-}$, & $A_L\;\;\Rightarrow $ & $\Delta \bar
u(x)/\bar u(x)$\\
 & & & & $\Delta d(x)/d(x)$\\[0.4cm]

$\bullet\; \vec p+ \vec p\to$ &
$\mu^+\mu^- +X\;\;\;$,&
$q +\bar q\to \mu^{+}+\mu^-$, & $A_{LL}\Rightarrow $ & $\Delta \bar
q(x)$\\
\end{tabular}
\end{center}

We do not provide the equations for one- and two-spin
asymmetries. The actual formulas certainly contain rather
complicated integrations and should be
analyzed with selection of appropriate kinematical limits.

2. Clarification  of the \underline{Strangeness content of a
nucleon} (and sea) requires the study of hyperon production
processes.
\begin{itemize}
\item $p+p\to \Lambda_\uparrow +X$, where, as always,
polarization of $\Lambda$ is studied through its decay process.
\item $\vec p+p\to \vec\Lambda+X$, measurement of the parameter
$D_{LL}$ in the fragmentation region at large $X_F$ values. The
estimates provide significant values for $D_{LL}$ close to 50\%.
\item $\vec p+\vec p\to\vec\Lambda +X$, measurement of the
three-spin correlation parameters $(l,l,l,0)$ would be important
for the study of hyperon production dynamics and the strangeness
content of the proton.
\end{itemize}

3. \underline{The transverse spin} can not be measured in DIS
because of the chiral invariance of eletromagnetic current.
\begin{center}
\begin{itemize}
\item $p_\uparrow+p_\uparrow \to\mu^+\mu^-+X,\;\;\;\;\;\;\;\;\;\;\;
q+\bar q\to \mu^+\mu^-,\;\;\;\;\;\;\;\;\;\;\;\;\; A_{NN}\Rightarrow h_1(x)$
\end{itemize}
\end{center}
The analysis based on operator product expansion shows that the
function $g_2(x, Q^2)$ in related to higher twists. Measuring
the function $h_1$ $(h_1=g_1?)$
one could judge on the role of higher twists at high energies as
well as non-perturbative effects in general.
\begin{center}
\begin{itemize}
\item
\begin{tabular}{llll}
$p_\uparrow+p_\uparrow\to Z^0+X$, &
$\,\,\,\,\,\,\,\,\,\,q+\bar q\to Z^0\to e^+e^-$, &
$\,\,\,\,\,\,\,\,\,\,\,A_{NN}\Rightarrow$ & $\Delta_\perp q(x)$\\
 & & & $\Delta_\perp \bar q(x)$\\
\end{tabular}
\item $p_\uparrow+p\to h^\pm+X$, $p_\perp$ --- dependence of the
asymmetry was not measured before. One may expect large $A_N$ at
large $p_\perp$ values. Experimental conclusion on the
prediction is an important test of $p$QCD and the role of
non-perturbative phenomena.
\end{itemize}
\end{center}

4. \underline{Elastic scattering} of polarized protons and
measurements of parameters $A_N$ and A$_{NN}$ is needed to
\begin{itemize}
\item confirm AGS effect at higher energies;
\item study spin structure of vacuum exchanges;
\item discriminate the models accounting non-perturbative dynamics.
\end{itemize}

5. Effects \underline{beyond the Standard Model} (compositeness,
SUSY). Both $A_L$ and $A_N$ measurements are useful in the
search for the compositeness. \\
$\bullet p+p\to \mbox{jet}+X$, $A_L$ is to be rather large at
$p_\perp\simeq 3-4$GeV\c. Compositeness should enlarge the
expected polarization effects. This deviation would arise from
the new interaction between quarks induced by their composite structure:
$$
L=L_{QCD}+\eta _0 \frac{g^2}{\Lambda^2_c}\bar q A q\bar q Aq,
$$
where $\Lambda_c$ is the scale of the compositeness of the order
of the binding energy for preons (1-2~TeV). The above said is
also true for production of direct photons or lepton pairs.

The asymmetries appearing in the production of SUSY particles in
the polarized hadron collisions will differ from the
corresponding asymmetries arising in the production of ordinary
particles. $A_{LL}$ is a relevant spin parameter. For the
processes where a pair of SUSY particles is produced, the
subprocess asymmetry $A_{LL}=-100\%$ for the case of massless
squarks and gluinos because of the helicity conservation. As a
result $A_{LL}$ should be negative (and have larger values)
contrary to the case of ordinary particles. So, one could expect
in the jet production process appearance of events with specific
behaviour of $A_{LL}$.

\section*{Inclusive Hadron Production}
Studies of one-spin transverse polarizations are of primarily
importance because of
\begin{itemize}
\item the existing experimental results (versus $p$QCD);
\item principal importance to clarify applicability of the "Born
formulas" of $p$QCD before going further to "the goals".
\end{itemize}

In this paragraph we discuss a possible origin of asymmetry on
pion production:
$$
p_\uparrow+p\to \pi^{\pm,o}+X.
$$
We use the scheme [16] which incorporates perturbative and
non-perturbative phases of QCD. It was already mentioned that
non-perturbative sector of QCD should provide the two important
phenomena: confinement and spontaneous breaking of chiral
symmetry with the relevant scales $\Lambda_{QCD}$ and
$\Lambda_{\chi}$. The chiral symmetry breaking results in
generation of $m_q\sim M$ and appearance of quark condensates
$<\bar \psi\psi>\not=0$.

A hadron is represented as a loosely bounded system of the
valence constituent quarks and quark condensate surrounding this
core. In this approach constituent quarks are extended objects.
They are described by their size and quark matter distribution:
$r_Q=\xi/m_Q,; d_Q(b)$.

The general form of the effective Lagrangian relevant for the
description of the non-perturbative phase of QCD
$$
L_{QCD}\to L_{eff}=L_\chi +L_I +l_c,
$$
where $L_\chi$ is responsible for the spontaneous chiral
symmetry breaking, $L_I$ --- for the constituent quarks
interaction, $L_c$ --- for the confinement. Both $L_I$ and $L_c$
do not affect the internal structure of constituent quarks. The
partonic structure of constituent quarks can be resolved in the
processes with large $Q^2$ values.

The particular form for $L_\chi$ is NJL model with 6-quark interaction:
\begin{eqnarray}
L_\chi&=&\bar\psi (i\gamma\cdot\hat\partial-\hat m)\psi+
\frac{1}{2}\sum_{a}G
[(\bar\psi\lambda_a\psi)^2+
(\bar\psi i\lambda_a\gamma_5\psi)^2]+\nonumber \\
&+&K[\mbox{det}\bar\psi_i(1-\gamma_5)\psi_j+
\mbox{det}\bar\psi_i(1+\gamma_5)].
\end{eqnarray}
Eq.(17) is a minimal effective Lagrangian which reflects some of
the basic properties of non-perturbative QCD.

The constituent quark masses are:
\begin{equation}
m_U=m_u-2G<0|\bar uu|0>
-2K<0|\bar dd|0><0|\bar ss|0>.
\end{equation}
Massive quarks appear as quasiparticles: as current quarks and
the surrounding clouds of $\bar qq$ --- pairs which consist of a
mixture of quarks of different flavors.

Quantum numbers of the constituent quarks are the same as the
quantum numbers of current quarks due to conservation of the
corresponding currents in QCD. Constituent quarks picture of a
hadron is consistent with the results for the proton spin
structure function $g_1(x)$ obtained in DIS. It is useful to
note that
$$
2<U|\bar s s|U>/<U|\bar u\bar u+\bar d d|U>\sim 0.15
$$
$$
<U|\bar u u|U>\gg<U|\bar dd|U>,\,\,\,
<U|\bar ss|U>.
$$

About the spin content of constituent quark. The axial anomaly
leads to compensation of valence quark helicity by helicities of
quarks from the cloud
\begin{equation}
<Q^i|A^j_{\mu_5}|Q^i>=\frac{1}{2}
\left (\delta^i_j-\frac{2}{3} c\right )s^i_\mu.
\end{equation}
Eq.(19) shows that constituent quark of any flavor contains a
sea of polarized current quarks of all other flavors.

Significant part of constituent quark spin should be associated
with the orbital angular momentum of quarks inside this
constituent quark (cloud quarks should
rotate coherently). What is the origin of this orbital angular
momentum? One can use an analogy between hadron physics and
superconductivity, where pairing  correlations induce particle
current around the anisotropy direction $\hat{\stackrel{\to}l}$.
Thus particle at the origin is surrounded by cloud of correlated
particles that rotate around
$\hat{\stackrel{\to}l}$. In our case the axis of anisotropy is
determined by the polarization vector of valence quark located
at the origin of constituent quark. The orbital angular momentum
$\vec L$ lies along $\hat{\stackrel{\to}l}$ and its value is
proportional to quark density $<Q|\bar q q|Q>$.

Thus, the spin of constituent quarks $J_U$ is determined by the
following sum:
\begin{equation}
J_U=\frac{1}{2}=J_{u_V}+J_{\{\bar qq\}}+<L_{\{\bar qq\}}>.
\end{equation}
Estimation of the orbital momentum contribution with account for
eq.(9) and $SU(6)$ model  leads to conclusion
$$
<L_{\{\bar qq\}}>\simeq 1/3.
$$
Orbital motion of quark matter inside constituent quark is the
origin of the asymmetries in inclusive production at moderate
and high transverse momenta. Such asymmetry will be significant at
$p_\perp \geq \Lambda_\chi \simeq 1$GeV. At high $p_\perp$
values we will have a parton picture for constituent  quark as a
cluster of non-interacting quarks which however should naturally
preserve their orbital momenta of the preceding
non-perturbative phase of QCD.

Without going in the details the asymmetry for the process
$$
h^\uparrow_1+h_2\to h_3+X
$$
is taking the form
\begin{equation}
A_N(s,x,p_\perp)=\sin (P_Q<L_{\{q\bar
q\}}>)d\sigma^{\mbox{\footnotesize hard}}/
\{d\sigma^{\mbox{\footnotesize hard}}
+d\sigma^{\mbox{\footnotesize soft}}\}.
\end{equation}
The sign and value of the asymmetry are determined by the
polarization $P_Q$ of the relevant constituent quark $Q$ inside
the hadron $h_1$ and the mean orbital momenta of cloud quarks:
$$
h_3=\pi^+,\,\,\,\,\, Q=U
$$
$$
h_3=\pi^-,\,\,\,\,\, Q=D.
$$
For instance, in $SU(6)$ model $P_U=2/3$ and $P_D=-1/3$. Fig.7
represents the model predictions for the asymmetries in
$p_\uparrow +p\to \pi^\pm+X$. $A_N$ has a weak energy dependence
and gets significant values starting from $p_\perp \simeq
1$GeV/c. The observed $p_\perp$-behaviour of asymmetries in
inclusive processes seems to confirm these conclusions.
Fig.8 shows the asymmetry in $\pi^0$-production.

Asymmetry reflects internal structure of the constituent quarks
and is proportional  to the orbital angular momentum of current
quarks inside the constituent quark. The significant asymmetries
appear beyond $p_\perp\geq 1$Gev/c, i.e. the scale where
the internal structure of constituent quark can be probed. The
proposed mechanism is also appropriate for description of hyperon
polarizations. It is worth noting here
that this idea
could be traced back to the model of rotating hadronic matter~[17].

The proposed mechanism for generation of the asymmetry differs
from the one when asymmetry appears at the level of
fragmentation function~[18]. In this connection we would like to
mention here the ALEPH result
on depolarization occurring during hadronization~[19].

\vspace*{0.4cm}
In conclusion we would like to underline that spin physics
program which can be conducted at RHIC is potentially very rich.
Polarization experiments always in the past provided particle
physics with unexpected new results and initiated searches for
deeper understanding of the fundamental dynamics.

I am pleased to thank S.M.Troshin for useful discussions and
suggestions under preparation of this talk.

\end{document}